\documentclass[%
reprint,
superscriptaddress,
amsmath,amssymb,
aps,
showkeys
]{revtex4-2}

\usepackage{graphicx}
\usepackage{dcolumn}
\usepackage{bm}
\usepackage{multirow}
\usepackage{physics}
\usepackage{hyperref}
\usepackage{xcolor}
\usepackage[normalem]{ulem}
\usepackage[linesnumbered,ruled,vlined]{algorithm2e}
\usepackage[caption=false]{subfig}
\usepackage{units}

\newtheorem{theorem}{Theorem}

\newcommand{\Cp}{{\mathcal{C}}} 
\newcommand{\Tp}{{\psi}} 

\begin{document}

\preprint{APS/123-QED}

\title{Schmidt quantum compressor}
\author{Israel F. Araujo}
\email{ifa@yonsei.ac.kr}
\affiliation{Department of Statistics and Data Science, Yonsei University, Seoul 03722, Republic of Korea}
\affiliation{Departamento de Eletrônica e Sistemas, Universidade Federal de Pernambuco, Recife, Pernambuco, 50.740-550, Brazil}
\affiliation{Centro de Informática, Universidade Federal de Pernambuco, Recife, Pernambuco, 50.740-560, Brazil}

\author{Hyeondo Oh}
\affiliation{Department of Statistics and Data Science, Yonsei University, Seoul 03722, Republic of Korea}

\author{Nayeli A. Rodr\'{i}guez-Briones}
\affiliation{Miller Institute for Basic Research in Science, University of California Berkeley, CA 94720, USA}
\affiliation{Atominstitut, Technische Universit{\"a}t Wien, Stadionallee 2, 1020 Vienna, Austria}

\author{Daniel K. Park}
\email{dkd.park@yonsei.ac.kr}
\affiliation{Department of Statistics and Data Science, Yonsei University, Seoul 03722, Republic of Korea}
\affiliation{Department of Applied Statistics, Yonsei University, Seoul 03722, Republic of Korea}

\begin{abstract}
This work introduces the Schmidt quantum compressor, an innovative approach to quantum data compression that leverages the principles of Schmidt decomposition to encode quantum information efficiently. In contrast to traditional variational quantum autoencoders, which depend on stochastic optimization and face challenges such as shot noise, barren plateaus, and non-convex optimization landscapes, our deterministic method substantially reduces the complexity and computational overhead of quantum data compression. We evaluate the performance of the compressor through numerical experiments, demonstrating its ability to achieve high fidelity in quantum state reconstruction compared to variational quantum algorithms. Furthermore, we demonstrate the practical utility of the Schmidt quantum compressor in one-class classification tasks.
\end{abstract}

\keywords{quantum computing, Schmidt decomposition, quantum compression, quantum autoencoder}

\maketitle

\section{Introduction}
\label{sec:introduction}

Quantum compression is a technique in quantum computing aimed at reducing the dimensionality of quantum information while preserving its essential characteristics~\cite{Nielsen_Chuang_2010,Watrous_2018}. This process involves encoding quantum states into a lower-dimensional space, allowing for more efficient storage and manipulation of quantum data. In this way, quantum compression can significantly optimize the use of quantum resources, making it a valuable tool in advancing quantum technologies.

Applications of quantum compression~\cite{pivoluska2022implementation,rozema2014quantum} span a variety of fields, including quantum simulation~\cite{fan2021efficient}, quantum communication~\cite{plesch2010efficient}, and distributed quantum computing~\cite{caleffi_distributed_2022, loke_distributed_2023}. By minimizing the quantum resources required, quantum compression enhances the feasibility and efficiency of quantum simulations, enables more effective quantum communication protocols by reducing the quantum data transmitted over networks, and facilitates distributed computation within quantum networks. Importantly, this technology is not limited to purely quantum data; it can also efficiently transmit classical data  through quantum networks by encoding it as a quantum state~\cite{mottonen2005, plesch2011, araujo2021divide, araujo2021configurable, araujo2024lowrank}, thereby expanding the range of information that can be compressed and shared. These applications highlight the potential of quantum compression to improve the way both quantum and classical information is processed and transmitted.

The quantum autoencoder (QAE)~\cite{Romero_2017,PhysRevApplied.15.054012,huang2020realization,MA2023110659,Lamata_2019} is a popular approach to achieving quantum compression by employing a two-part mechanism: an encoder and a decoder. The compressor maps the high-dimensional input quantum states into a lower-dimensional latent space, effectively compressing the information. This is followed by a decompressor, which attempts to reconstruct the original quantum state from the compressed version. The success of a QAE can be measured by the fidelity between the original and reconstructed states, which indicates how accurately the compression and subsequent decompression preserve the quantum information. A QAE can be trained to maximize compression efficiency by optimizing the unitary transformations that define the encoding and decoding processes, highlighting its significant contribution to the field of quantum data compression.

However, QAEs face several challenges that limit their efficiency in compressing quantum data, primarily due to their foundation on variational quantum circuits.
These variational circuits rely on fine-tuning their parameters through optimization techniques. This approach is inherently stochastic, as both the objective function and its gradient with respect to the optimization parameters are formulated in terms of expectation values. Consequently, the optimization process is hindered by shot noise, necessitating numerous repetitions of the circuit to accurately determine gradient directions, while also limiting scalability by increasing the likelihood of encountering barren plateaus~\cite{mcclean_barren_2018,Cerezo2021bp, larocca2024review}. Moreover, the non-convex nature of the optimization landscape can cause training trajectories to easily become trapped in local minima~\cite{PhysRevLett.127.120502,Anschuetz2022}.

Additionally, identifying the optimal circuit configuration becomes increasingly challenging~\cite{hur2023neural,Lourens2023_qnas}.
The circuit configuration consists of two parts: an embedding, which maps classical data into a Hilbert space, and an ansatz, a parametrized circuit designed to approximate the solution to a given problem. Choosing the right embedding and ansatz for the circuit configuration demands a balance between the circuit's ability to represent complex states (expressibility~\cite{schuld2021effect}) and the ease of optimization (trainability~\cite{thanasilp2022exponential}). The embedding must efficiently translate classical data into a quantum context, ensuring information is encoded without loss; however, when working with quantum data, this step is unnecessary~\cite{10.1038/s43588-022-00311-3}. On the ansatz side, the architecture of the variational circuit should be sufficiently sophisticated to represent the dataset's complexity without becoming unmanageable. An ansatz that is too simplistic risks failing to model the data adequately, while one that is too elaborate may induce barren plateaus~\cite{mcclean_barren_2018, holmes_2022} in the optimization landscape. Moreover, introducing more gates into the circuit tends to increase error propagation in noisy intermediate-scale quantum (NISQ) devices, potentially corrupting the quantum state. Given these considerations, the careful selection of both embedding and ansatz is critical, as it profoundly impacts the effectiveness of variational quantum computing methods.

Furthermore, the variational nature of these circuits requires a significant amount of classical computational resources for optimization, which can be time-consuming. This is particularly challenging when dealing with complex quantum systems or scaling the QAE for larger quantum datasets. The trade-off between the adaptability of variational circuits to different quantum compression tasks and the computational overhead introduced by the optimization process, along with the presence of noise, requires robust error mitigation and optimization strategies in developing QAEs.

In this work, we introduce a deterministic approach to quantum data compression to overcome these challenges inherent in variational quantum algorithms. Our approach leverages the Schmidt decomposition to transform quantum states into a form that can be efficiently compressed and later reconstructed. By constructing quantum circuits based on the Schmidt coefficients and basis vectors, this technique eliminates the need for parameter optimization and reduces susceptibility to noise and local minima. As a result, it bypasses the stochastic limitations of variational methods.

The paper is organized as follows.
Section~\ref{sec:Schmidt quantum state prep} describes a state preparation method that leverages the Schmidt decomposition, from which the underlying principle of our proposed approach is derived. Understanding this state preparation is essential for comprehending our new method.
Section~\ref{sec:Schmidt compression protocol} introduces the Schmidt Quantum Compression (SQC) protocol, which represents our main contribution in this work. Section~\ref{sec:experiments} showcases numerical experiments to evaluate the performance of the Schmidt quantum compressor and its comparison with the QAE. Subsequently, Section~\ref{sec:classification} demonstrates the performance of SQC in a classification problem, showing its comparability to the QAE. Section~\ref{sec:optimizations} discusses further optimizations that enhance the reconstruction fidelity of SQC.
Finally, Section~\ref{sec:conc} concludes the paper and discusses future directions.

\section{Schmidt Quantum State Preparation}
\label{sec:Schmidt quantum state prep}

\begin{figure}[htbp]
    \centering
    \includegraphics[width=1.0\columnwidth]{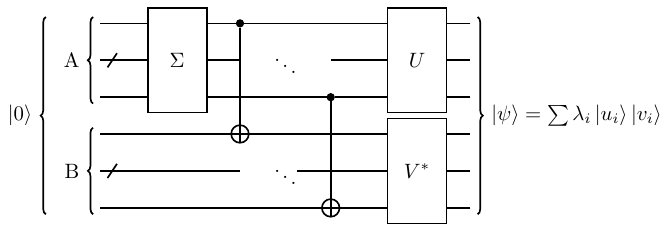}
    \caption{Schmidt Quantum State Preparation: quantum circuit overview to prepare $\ket{\psi}$ starting with $\ket{0}$. The operator $\Sigma$ encodes the singular values of the reshaped state vector of $\ket{\psi}$ as the amplitude probabilities of subsystem A (the first half of the quantum register). A sequence of CNOT gates then entangles subsystems A and B. The operators $U$ and $V$, derived from the state vector $\ket{\psi}$, transform the computational basis of subsystems A and B to align with the Schmidt basis.}
    \label{fig:schmidt}
\end{figure}

\begin{figure*}[htbp]
    \centering
    \includegraphics[width=1.0\columnwidth]{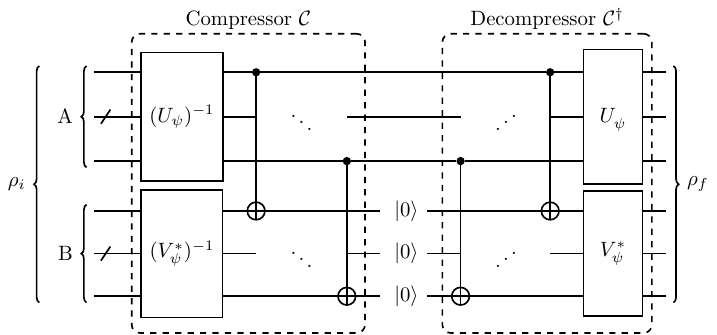}
    \caption{Schmidt Quantum Compression (SQC) protocol: quantum circuit overview. The operators $U_\Tp$ and $V_\Tp$ are derived from the typical state $\ket{\Tp}$. They transform the basis of subsystems A and B to align with the computational basis. A sequence of CNOT gates then disentangles the two subsystems. When $\rho_i$ is identical to $\ket{\Tp}$, the separation is perfect, and subsystem A carries the singular values of $\rho_i$. Consequently, the decompressor $\mathcal{C}^\dagger$ can faithfully reproduce $\rho_i$ without loss. For $\rho_i$ differing from $\ket{\Tp}$, the separation is approximate, with the quality of approximation $\rho_f$ being dependent on the proximity between $\rho_i$ and $\ket{\Tp}$.}
    \label{fig:overview}
\end{figure*}

Quantum state preparation refers to the process of mapping classical data to a quantum state, which serves as input for quantum algorithms in machine learning and other tasks~\cite{rebentrost2018quantum, wossnig2018quantum, childs2017quantum, lloyd2014quantum, blank2020quantum, levine2019quantum, benedetti2019parameterized, schuld2018supervised, schuld2017implementing, stoudenmire2017supervised, lloyd2013quantum, park2019circuit, giovannetti2008quantum, trugenberger2002quantum, ventura2000quantum, trugenberger2001probabilistic}.

Several methods exist for quantum state preparation, with some of the most common types being Amplitude Encoding~\cite{mottonen2004quantum, mottonen2005, bergholm2005quantum, veras2020circuit, schuld2020circuit, larose2020robust, araujo2021divide, araujo2021configurable}, Qubit Encoding (also known as Angle Encoding)~\cite{grant2018hierarchical}, and Hamiltonian Encoding~\cite{cade2020strategies, schuld2018supervised}.

This section provides a review of the Schmidt Quantum State Preparation~\cite{plesch2011, araujo2024lowrank}, a method classified under Amplitude Encoding. The primary objective of the Schmidt Quantum State Preparation is to establish a deterministic algorithm for state preparation that designs circuits whose depth is determined by the level of entanglement~\cite{araujo2024lowrank}. This state preparation forms the foundation for the protocol introduced by this work in Section~\ref{sec:Schmidt compression protocol}.

The Schmidt decomposition of a given bipartite state $\ket{\Tp}$ in the Hilbert space $\mathcal{H}_\textsc{a} \otimes \mathcal{H}_\textsc{b}$ can be written as
\begin{equation} \label{eq:schmidt}
    \ket{\Tp} = \sum_{i=1}^{k} \lambda_i \ket{u_i}_\textsc{a} \ket{v_i}_\textsc{b},
\end{equation}
where $\lambda_{i}$ are the Schmidt coefficients associated with this decomposition, and $\ket{u_i}_\textsc{a}$ and $\ket{v_i}_\textsc{b}$ correspond to the Schmidt basis vectors for subsystem A and B, respectively.
The quantum state $\ket{\Tp}$ is reshaped into a matrix $M_\Tp$ such that the Schmidt basis vectors of subsystem A correspond to the rows of $M_\Tp$, and the Schmidt basis vectors of subsystem B correspond to the columns of $M_\Tp$.
Performing Singular Value Decomposition (SVD) on this matrix $M_\psi$ yields:
\begin{equation} \label{eq:svd}
    M_\psi = U \Sigma V^\dagger,
\end{equation}
where the columns of $U$ (left singular vectors) correspond to the Schmidt basis vectors $\ket{u_i}_\textsc{a}$; the columns of $V$ (right singular vectors) correspond to the Schmidt basis vectors $\ket{v_i}_\textsc{b}$; and the diagonal elements of $\Sigma$ are the singular values, which are equivalent to the Schmidt coefficients $\lambda_i$. Thus, performing SVD on the reshaped matrix $M_\psi$ provides both the Schmidt basis vectors and the Schmidt coefficients for the bipartite state.

The SVD operation is reversible, which means that from the unitary components $U \Sigma V^\dagger$, one can reconstruct the original state $\ket{\psi}$ from $\ket{0}$. The equations below illustrate this reverse process:
\begin{equation}
    \ket{\lambda} = \Sigma \ket{0}_\textsc{a} = \sum_i \lambda_i \ket{i}_\textsc{a} = \ket{\lambda}_\textsc{a}
\end{equation}
\begin{equation} \label{eq:schmidt_measure}
    \left( \prod_{i=1}^{m} \text{CNOT}_{i}\right) \ket{\lambda}_\textsc{a} \ket{0}_\textsc{b} = \sum_i \lambda_i \ket{i}_\textsc{a} \ket{i}_\textsc{b}
\end{equation}
\begin{align}
    &\left(U\otimes V^*\right)\sum_i \lambda_i \ket{i}_\textsc{a} \ket{i}_\textsc{b} \nonumber \\
    &= \left[\Bigg(\sum_t \ket{u_t}\bra{t}\Bigg)_\textsc{a} \otimes \Bigg(\sum_j \ket{v_j}\bra{j}\Bigg)_\textsc{b}\right] \sum_i \lambda_i \ket{i}_\textsc{a} \ket{i}_\textsc{b} \nonumber \\
    &= \Bigg(\sum_j \ket{u_j}_\textsc{a}\ket{v_j}_\textsc{b} \bra{j}_\textsc{a}\bra{j}_\textsc{b}\Bigg) \sum_i \lambda_i \ket{i}_\textsc{a} \ket{i}_\textsc{b} \nonumber \\
    &= \sum_i \lambda_i \ket{u_i}\ket{v_i},
\end{align}
where $\text{CNOT}_{i}$ denotes controlled-NOT gates controlled by the $i$th qubit of subsystem A and targeting the $i$th qubit of subsystem B, and $m = \lceil \log_2(k)\rceil$ represents the Schmidt measure.

This entire procedure, encompassing the classical SVD (as described in Equation~\eqref{eq:svd}), the construction of the corresponding quantum circuit (depicted in Figure~\ref{fig:schmidt}), and the circuit's initialization to ultimately generate the targeted quantum state, is collectively referred to as the Schmidt Quantum State Preparation.

It is important to note that the partitioning of qubits into subsystems does not need to be continuous or uniform in size, as illustrated in Figure~\ref{fig:schmidt}. By strategically rearranging the qubits, it is possible to identify a bipartition where the bond dimension---defined as the maximum Schmidt measure $m$, or equivalently, the number of nonzero Schmidt coefficients $\lambda_i$ (see Equations~\eqref{eq:schmidt} and~\eqref{eq:schmidt_measure})---between the subsystems is minimized. This minimization indicates weak entanglement between the subsystems, which arises due to a reduction in the number of CNOT gates, as these gates have no effect when the corresponding Schmidt coefficients are zero and can be omitted. 

This configuration enables the state preparation protocol to design highly efficient circuits by employing isometries to optimize the circuit structure~\cite{Iten2016}. Since the columns of $U$ and $V^*$ corresponding to the zeroed Schmidt coefficients are unnecessary, they can be omitted. Consequently, the operations are implemented as $2^m$-to-$2^{n_j}$ isometries rather than full $2^{n_j} \times 2^{n_j}$ unitaries, significantly reducing the number of native gates required for the decomposition~\cite{araujo2024lowrank}.  Here, $j$ denotes the subsystem A or B, and $n_j \le m$ represents the number of qubits in the corresponding subsystem.

Finding a configuration with a minimal bond dimension further permits a low-error approximation of the quantum state by dropping the last $k-r$ Schmidt coefficients, where $r<k$ represents the number of retained coefficients. This approach allows for the partial ($r>1$) or complete ($r=1$) decoupling of qubit subsets, where the fidelity loss $\displaystyle l(r, \ket{\psi}):= (1-\|\langle \psi|\psi^{(r)}\rangle\|^2)=\sum_{i=r+1}^k \|\lambda_i\|^2$ is proportional to the magnitude of the dropped Schmidt coefficients.

This strategy, known as the low-rank approximation~\cite{araujo2024lowrank}, enables the design of less complex circuits by reducing the requirement for CNOT gates to entangle subsystems A and B from $\lceil \log_2(k)\rceil$ to $\lceil \log_2(r)\rceil$ gates. Consequently, it leverages the aforementioned isometries to achieve this simplification, allowing for more resource-efficient quantum computations. Further optimization can be achieved for sparse isometries~\cite{Malvetti2021quantumcircuits}.

\section{Schmidt quantum compression protocol}
\label{sec:Schmidt compression protocol}
The goal of our SQC protocol, as illustrated in Figure~\ref{fig:overview} and detailed in Algorithm~\ref{alg:compressor}, is to design a quantum data compression scheme---based on either a classical dataset or samples from a quantum data source---that maximizes the fidelity between the input state $\rho_i$ and the final state $\rho_f$.
The core of the method is to identify a state $\ket{\Tp}$, referred to as the \textit{typical state}, that minimizes the distance to a given set of sample states $\{\ket{x_i}\}$ (see Section~\ref{sec:avg_state}).

\subsection{Protocol principle}

Similarly to Section~\ref{sec:Schmidt quantum state prep}, if the typical state $\ket{\Tp}$ is reshaped into a matrix form $M_\psi$, and an SVD is performed on this matrix, the result is:
\begin{equation} \label{eq:svd_typical}
    M_\Tp = U_\Tp \Sigma_\Tp V_\Tp^\dagger.
\end{equation}

\begin{figure}[htbp]
    \centering
    \includegraphics[width=0.71\columnwidth]{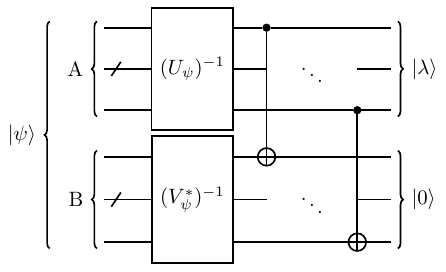}
    \caption{The figure showcases a compressor circuit adapted from the Schmidt Quantum State Preparation. In this adaptation, the original circuit is inverted, and the unitary component $\Sigma$ is omitted. Upon application to a typical state $\ket{\Tp}$, the circuit produces a new state $\ket{\lambda}$, which encodes the singular values formerly attributed to $\Sigma_\Tp$ within subsystem A. After this transformation, subsystems A and B become fully disentangled, leaving subsystem B in the state $\ket{0}$.}
    \label{fig:compressor}
\end{figure}

The compression unitary $\Cp$, depicted in Figure~\ref{fig:overview}, is formulated from Equation~\eqref{eq:svd_typical} and is derived from the Schmidt state preparation circuit (see Section~\ref{sec:Schmidt quantum state prep}). Observing that the first operator in the Schmidt circuit (see Figure~\ref{fig:schmidt}) is a state preparation mechanism that encodes the singular values into the amplitudes of subsystem A, it is feasible to reverse this circuit and eliminate the associated operator $\Sigma$ to generate the state $\ket{\lambda}$ (see Figure~\ref{fig:compressor}). With $\Cp$ as the compression unitary, $\Cp^\dagger$ serves as its corresponding decompression unitary.

\begin{figure}[htbp]
    \centering
    \subfloat[]{ \label{fig:overview_exact}
        \includegraphics[width=0.81\columnwidth]{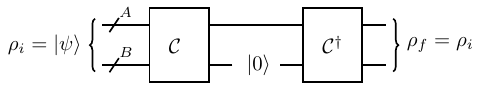}
    } \\
    \subfloat[]{ \label{fig:overview_approx}
        \includegraphics[width=0.81\columnwidth]{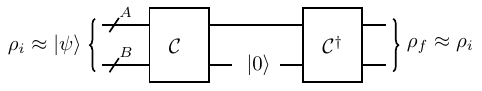}
    }
    \caption{Operational dynamics of the complete circuit on various states. (a) Scenario where the input state precisely matches the typical state, enabling a lossless recovery of the input state. (b) Scenario where the input state is in close proximity to the typical state, yielding a high-fidelity approximation of the input state, although not an exact recovery.}
\end{figure}

When $\Cp$ acts on the typical state $\ket{\Tp}$, it precisely produces $\ket{\lambda}$ in subsystem A and $\ket{0}$ in subsystem B, as shown in Equation~\eqref{eq:decomp_typical}. In this configuration, $\Cp^\dagger$ can perfectly reconstruct $\ket{\Tp}$, as illustrated in Figure~\ref{fig:overview_exact}.

\begin{equation}
    (U_\Tp \otimes V_\Tp^*)^{-1} \ket{\Tp} = \sum_i \lambda_i \ket{i}_\textsc{a} \ket{i}_\textsc{b}
\end{equation}
\begin{equation} \label{eq:decomp_typical}
    \left( \prod_{i=1}^{m} \text{CNOT}_{i} \right)\sum_i \lambda_i \ket{i}_\textsc{a} \ket{i}_\textsc{b} = \sum_i \lambda_i \ket{i}_\textsc{a} \ket{0}_\textsc{b}
\end{equation}

Now, consider a state $\ket{x_i}$ that is close to $\ket{\Tp}$. We are interested in how $\ket{x_i}$ projects onto the known Schmidt basis vectors of $\ket{\Tp}$. The projection is expressed as:
\begin{equation}
    \alpha_i = \bra{u_i}_\textsc{a} \bra{v_i}_\textsc{b} \ket{x_i}
\end{equation}
The projections translate in terms of closeness in the following way:
\begin{itemize}
    \item \textbf{Exact match.} If $\ket{x_i} = \ket{\Tp}$, the coefficients $\alpha_i$ will exactly match  with the Schmidt coefficients $\lambda_i$.
    \item \textbf{Close but not exact.} If $\ket{x_i}$ is near $\ket{\Tp }$ (in terms of a high fidelity between them), the coefficients $\alpha_i$ will be close to $\lambda_i$, but with some deviations. The magnitude and pattern of these deviations provide insights into how $\ket{x_i}$ deviates from $\ket{\Tp}$.
    \item \textbf{Orthogonal or unrelated.} If $\ket{x_i}$ is orthogonal to or largely unrelated to $\ket{\Tp}$ in terms of certain Schmidt modes, the corresponding $\alpha_i$ values will be small or even zero.
\end{itemize}
The entire set $\{ \alpha_i \}$ can be seen as a signature or fingerprint of how $\ket{x_i}$ projects onto the Schmidt basis of $\ket{\Tp}$.

\begin{algorithm}[htbp]
    \SetKwInOut{Input}{input}\SetKwInOut{Output}{output}
    \Input{Typical state vector $\ket{\Tp}$ with $2^n$ amplitudes.}
    \Input{A partition block $b_\textsc{a}$ of $n_\textsc{a}$ qubits ($0 < n_\textsc{a} < n)$.}
    \Output{Compressor quantum circuit.}
    \BlankLine
    Reshape $\ket{\Tp}$ into a matrix $M_\Tp$ according to the sizes of the partition blocks ($n_\textsc{a}$ lines and $n-n_\textsc{a}$ columns) and $b_\textsc{a}$ configuration (logical qubits swap). \label{st:reshape} \\
    Decompose $M_\Tp$ using \textit{SVD} (store unitaries $U_\Tp$, $V^\dagger_\Tp$, and $\Sigma_\Tp$). \label{st:start_plesch} \\
    Set rank equal to the number of non-zero elements of the $\Sigma_\Tp$ diagonal $\sigma$ ($\text{rank}=\text{count}(\sigma \ne 0)$). \\
    Set Schmidt measure $m=\lceil \log_2(\text{rank}) \rceil$. \\
    If $m < n_\textsc{a}$, set the number of columns of $U_\Tp$ and $(V^\dagger_\Tp)^T=V^*_\Tp$ to $2^m$, otherwise continue. \\
    Create a quantum circuit with $n=\log_2(\text{length}(\ket{\Tp}))$ qubits. \\
    Encode $(U_\Tp)^{-1}$ on qubits $q \in b_\textsc{a}$ using an appropriate algorithm according to its dimension (vector, isometry, or unitary). \\
    Encode $(V^*_\Tp)^{-1}$ on qubits $q \notin b_\textsc{a}$ using an algorithm according to its dimension (vector, isometry or unitary). \\
    Perform $m$ \textit{CNOT} gates between control qubit $q_c \in b_\textsc{a}$ and target $q_t \notin b_\textsc{a}$. \\
    Output the compressor quantum circuit.
    \caption{Schmidt quantum compressor}
    \label{alg:compressor}
\end{algorithm}

\subsection{Compressor action}

The protocol is designed to encode $n$ qubits into $n_\textsc{a}$ qubits, referred to as the \textit{latent space}. The remaining $n_\textsc{b}=n-n_\textsc{a}$ qubits are termed \textit{trash qubits}. We designate the latent space as subsystem A and the trash space as subsystem B. Thus, $\mathcal{H}_\textsc{ab}$ refers to the entire original Hilbert space, where $\mathcal{H}_\textsc{a}$ corresponds to the latent space, and $\mathcal{H}_\textsc{b}$ to the trash space. Given an input state represented by $\rho_i = \ket{x_i}\bra{x_i}$, we can apply a unitary compressor $\Cp$, constructed from the typical state $\ket{\Tp}$ (see Figures~\ref{fig:overview} and~\ref{fig:compressor}), defined as
\begin{equation}
    \Cp = \left( \prod_{i=1}^{m} \text{CNOT}_{i} \right) \left( U_\Tp \otimes V_\Tp^* \right)^{-1},
\end{equation}
to transform the input state into: 
\begin{equation}
    \rho_c = \Cp\rho_i\Cp^\dagger.
\end{equation}
From this, the states associated with the latent and trash spaces can be derived as:
\begin{align}
    \rho_l &= \mathrm{Tr}_\textsc{b}(\rho_c), \\
    \rho_t &= \mathrm{Tr}_\textsc{a}(\rho_c).
\end{align}
To decompress the latent state $\rho_l$, the inverse operation $\Cp^\dagger$ is applied. The final state $\rho_f$ is obtained through the full quantum circuit (Figure~\ref{fig:overview}) and is expressed as
\begin{align}
    \rho_f &= \Cp^{\dagger} \left[\Tr_\textsc{b}\left(\Cp \rho_i \Cp^{\dagger}\right) \otimes \rho_\text{ref}\right]\Cp \\ &= \Cp^\dagger \left(\rho_l \otimes \rho_\mathrm{ref}\right) \Cp,
\end{align}
where $\rho_\mathrm{ref}$ is the reference state, typically chosen as $(\ket{0}\bra{0})^{\otimes n_\textsc{b}}$.

\subsection{Compressor complexity}

The complexity of the SQC algorithm, being based on the Schmidt Quantum State Preparation, aligns closely with the results presented in~\cite{araujo2024lowrank}. Let $\ket{\psi}$ be an $n$-qubit quantum state with a Schmidt decomposition, where subsystem $\mathcal{H}_\textsc{b}$ consists of $1 \leq n_\textsc{b} \leq \lfloor n/2 \rfloor$ qubits. Without loss of generality (since $n_\textsc{b}$ can always be swapped with $n_\textsc{a}$ if $n_\textsc{a} < n_\textsc{b}$), and given that a unitary decomposition typically requires $\nicefrac{23}{48}(2^{2s}) - \nicefrac{3}{2}(2^s) + \nicefrac{4}{3}$ CNOT gates for $s$ qubits~\cite{shende2006synthesis}, while the decomposition of general isometries from $m$ qubits to $s > m$ qubits requires $2^{m+s} - \nicefrac{1}{24}(2^s) + O(s^2)2^m$ CNOT gates~\cite{Iten2016}, the SQC algorithm constructs quantum circuits where the number of CNOT gates is given by:

\begin{footnotesize}
\begin{itemize}

\item $0 \le m < n_\textsc{b}$
\begin{equation*} \label{eq:iso_cnots}
    \underbrace{ 2^{m+n_\textsc{a}}-\frac{1}{24}2^{n_\textsc{a}} }_\text{subsystem A (isometry)} + \underbrace{ 2^{m+n_\textsc{b}}-\frac{1}{24}2^{n_\textsc{b}} }_\text{subsystem B (isometry)} + \underbrace{ m \vphantom{\frac{1}{24}} }_\text{CNOTs}
\end{equation*}

\item $m=n_\textsc{b}$ and $n_\textsc{b}<n_\textsc{a}$
\begin{equation*} \label{eq:uni_iso_cnots}
    \underbrace{ 2^{n}-\frac{1}{24}2^{n_\textsc{a}} }_\text{subsystem A (isometry)} + \underbrace{ \frac{23}{48}2^{2n_\textsc{b}} - \frac{3}{2}2^{n_\textsc{b}} + \frac{4}{3} }_\text{subsystem B (unitary)} + \underbrace{ n_\textsc{b} \vphantom{\frac{1}{24}} }_\text{CNOTs}
\end{equation*}

\item $m=n_\textsc{b}$ and $n_\textsc{b}=n_\textsc{a}$
\begin{equation*} \label{eq:uni_cnots}
    \underbrace{2 \left( \frac{23}{48}2^{n} - \frac{3}{2}2^{n_\textsc{b}} + \frac{4}{3} \right) }_\text{subsystems A and B (unitaries)} + \underbrace{ n_\textsc{b} \vphantom{\frac{1}{24}} }_\text{CNOTs}
\end{equation*}
\end{itemize}
\end{footnotesize}
These equations are bounded by the results of Theorem~\ref{thm:sqc}.

\begin{theorem}[Schmidt Quantum Compressor.]\label{thm:sqc}
Given Equation~\eqref{eq:schmidt} with the Schmidt measure $m = \lceil \log_2(k)\rceil$, the circuit produced by the SQC protocol has a complexity described as follows:
\begin{table}[htbp]
    \centering
    \begin{tabular}{c|c}\hline
        Condition & CNOT Count \\ \hline
        $0\le m <n_\textsc{b}$ & $O(2^{m+n_\textsc{a}})$ \\ \hline
        $m=n_\textsc{b}$ & $O(2^{n})$ \\ \hline
    \end{tabular}
\end{table}
\end{theorem}

In the context of classical preprocessing and scalability, the execution of singular value decomposition (SVD) on the typical state, as outlined in Equation~\eqref{eq:svd_typical}, requires approximately $\sim 2^n$ bytes of memory for states involving $n$ qubits. Additionally, the time complexity for this operation scales as $O(2^{3n/2})$~\cite{KOGBETLIANTZ_1955,forsythe_1960}. This process demands considerable computational resources as the number of qubits increases. To overcome this limitation, Randomized SVD can be employed to produce a low-rank approximation with a complexity of $O(2^n)$~\cite{ApproximateMatrixDecompositions2011, RandomizedSVD2021}. However, it is important to note that this calculation is a one-time requirement after the typical state is identified, and its results can be leveraged for any further assessments of the quantum circuit. Additionally, the complexity of the SVD process does not depend on the dataset size.

Identifying the bipartition with the minimum bond dimension---as discussed in Section~\ref{sec:Schmidt quantum state prep}---involves a search whose complexity is determined by the size of the smaller block, $n_\textsc{b}$, and grows according to the binomial coefficient $\displaystyle\binom{n}{n_\textsc{b}}$. Consequently, in the worst-case scenario, i.e., when the bipartition blocks have the same size,  this search reaches a complexity of $O(2^n)$. Calculating the bond dimension during each search iteration can be approached either exactly, which demands considerable computational resources for larger systems, or through heuristic methods that potentially offer more efficiency by approximating the result.

When dealing with quantum data, the SQC must extract samples from the quantum source to produce the typical state, a process often achieved through quantum tomography. However, due to the exponential cost of tomography, this approach is frequently impractical. A more feasible alternative is to use classical shadow approximation~\cite{huang2020}. Nevertheless, when working with a large number of qubits (greater than 50), extracting samples presents a significant challenge for the SQC  due to the size of the states involved. From a practical standpoint, this limitation can be addressed by leveraging the fact that, in current quantum devices, large quantum states can often be expressed as a tensor product of smaller states. This is achievable by considering the limited qubit connectivity and the inherent noise in current devices, which permits the use of error approximation techniques~\cite{araujo2024lowrank,PhysRevX.10.041038}.

It can be argued that when working with quantum data in a large quantum space (over 50 qubits), QAE may offer an advantage over the SQC. This is because the parameterized circuit in QAE can be designed with a low depth, which, while reducing the model's expressibility, may still produce a satisfactory approximation—though success is not guaranteed. This approach also avoids the need to generate large data samples to determine the typical state and to compute the SVD of the corresponding matrix. However, this claim is primarily based on small-scale quantum machine learning (QML) experiments. Demonstrating that quantum models perform well in these limited scenarios does not provide sufficient evidence that QML will be effective for realistic problem sizes~\cite{bowles2024betterclassicalsubtleart}. This caution also applies to QAEs, which operate on similar principles as QML. Moreover, this claim fails to consider that the SQC can use low-rank approximations to reduce the depth of the circuit~\cite{araujo2024lowrank}, making it competitive with QAE in terms of circuit complexity.

\subsection{Average state as the typical state}
\label{sec:avg_state}

A key objective of the compressor is to maximize the fidelity between the input and final states after the reconstruction. The fidelity we aim to maximize is
\begin{equation} \label{eq:fidelity}
    F\left(\ket{x_i}, \rho_f\right) = \bra{x_i}\rho_f\ket{x_i}.
\end{equation}
The fidelity gives a measure of how close $\rho_f$ (the recovered state) is to the original state $\rho_i$ (the input state). A value of 1 indicates perfect agreement, while a value less than 1 signifies some degree of discrepancy. Thus, the fidelity value strongly depends on how well the full process preserves the quantum information of $\ket{x_i}$. If the process is not perfectly unitary or if information is lost (due to the partial trace), the fidelity will be less than 1.

From Equation~\ref{eq:fidelity}, the fidelity between the input and the recovered states is determined by the closeness between the typical state $\ket{\Tp}$ and $\ket{x_i}$.
That is, if the input state is close to the typical state, $\ket{x_i}\approx \tilde{\ket{\Tp}}\approx \ket{\Tp}$, leads to the following expression for fidelity:
\begin{align*}
    F(\ket{x_i}, \rho_f) 
    &= \bra{x_i} \Cp^{\dagger} (\rho_l \otimes \rho_\mathrm{ref}) \Cp  \ket{x_i} \\
    &= \braket{x_i}{\tilde{\Tp}} \braket{\tilde{\Tp}}{x_i} \\
    &\approx 1,
\end{align*}
where $\rho_l$ is close to the typical state singular values, leading to a recovered state that is close to $\ket{\Tp}$, and, consequently,  close to $\ket{x_i}$. This result implies that, to maximize fidelity, one must minimize the distance between $\ket{x_i}$ and $\ket{\Tp}$.

Thus, our goal is to find a typical state $\ket{\Tp}$ that minimizes the distance between the entire sample set $\{\ket{x_i}\}$ and $\ket{\Tp}$, thereby maximizing the fidelity of the entire dataset to 1.

To analyze this, we aim to minimize the distance between the input state and the typical state by introducing the $L_2$ distance as follows:
\begin{align}
\displaystyle
    l_i(\ket{x_i}, \ket{\Tp}) :&= \| \ket{x_i} - \ket{\Tp} \|_2^2 \\
    &= \braket{x_i}{x_i} - 2\text{Re}\braket{x_i}{\Tp} + \braket{\Tp}{\Tp}.
\end{align}

Here, $l_i$ represents the squared $L_2$ norm of the difference between the input state $\ket{x_i}$ and the typical state $\ket{\Tp}$. Considering the entire sample set, we define the total distance $D(\ket{\Tp})$ as the sum of all individual distances:
\begin{equation}
    D(\ket{\Tp}) = \sum_{i=1}^{M} l_i(\ket{x_i}, \ket{\psi}).
\end{equation}

To find the typical state that minimizes the total distance, we set the derivative of $D(\ket{\Tp})$ with respect to $\ket{\Tp}$ to zero. In our study, we utilized a classical image dataset, where the pixel values are inherently real and non-negative. Consequently, the states $\ket{x_i}$ are real, non-negative, and normalized vectors. The derivative is done taking into account that the values are real. In addition, the absence of negative values is important because it could result in an average state of zero length (and thus no direction). These characteristics enable the average state to serve as the optimal typical state.
\begin{equation} \label{eq:derivative}
    \frac{d}{d\ket{\Tp}}D(\ket{\Tp}) = -2\sum_{i=1}^{M} \ket{x_i} + 2M\ket{\Tp} = 0.
\end{equation}

Solving Equation~\eqref{eq:derivative} yields the typical state as the average of the input states:
\begin{equation}
\ket{\Tp} = \frac{\sum_{i=1}^{M} \ket{x_i}}{M}.
\end{equation}
The normalization might not be preserved after summing up the state vectors, so $\ket{\Tp}$ may need to be renormalized afterward---normalization does not change the direction of a vector, only its magnitude.

One of the primary advantages of the average is simplicity. Calculating the average state is computationally straightforward. Regarding stability, a small change in the dataset (such as adding or removing a few states) might not drastically affect the average, making it a relatively stable measure. Suppose a single additional value, $\ket{x_{M+1}}$, is added to the dataset. The new average $\ket{\Tp'}$ becomes:  
\begin{equation}
\ket{\Tp'} = \frac{1}{M+1} \left( \sum_{i=1}^M \ket{x_i} + \ket{x_{M+1}} \right)
\end{equation}
Expanding in terms of the original average $\ket{\Tp}$:  
\begin{equation}
\ket{\Tp'} = \frac{M}{M+1} \ket{\Tp} + \frac{1}{M+1} \ket{x_{M+1}}
\end{equation}
This shows that the new average is a weighted combination of the old average and the new value $\ket{x_{M+1}}$, ensuring stability for large $M$.

\begin{figure*}[ht]
    \centering
    \includegraphics[width=1.0\textwidth]{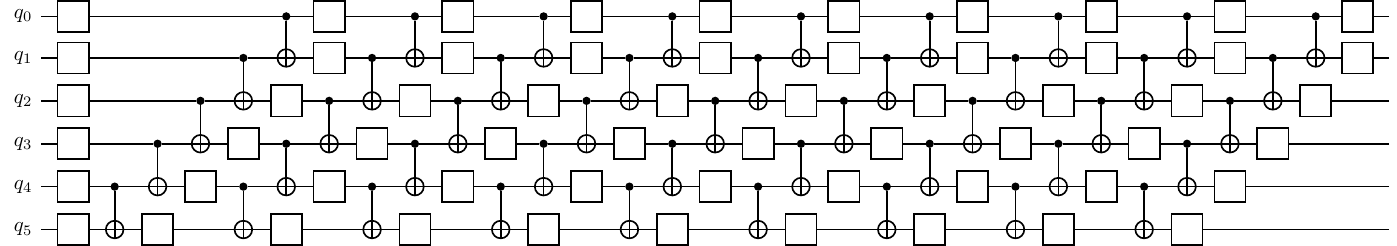}
    \caption{The diagram illustrates the ansatz used as the compression unitary within QAE experiments, designed to ensure competitive comparison based on circuit complexity. This particular ansatz comprises 45 CNOT gates, in contrast to the SQC unitaries, which feature between 40 and 43 CNOT gates. The variability in the number of CNOTs for SQC unitaries reflects the algorithm's adaptability to the entanglement level of the typical state, with the precise count fluctuating according to the dataset class. In addition to the CNOT gates, the circuit contains sixty single-qubit unitary gates, each implementing a parameterized $R_y(\theta_i)$ rotation.}
    \label{fig:ansatz}
\end{figure*}

\section{Experiments}
\label{sec:experiments}

\subsection{Comparison with quantum autoencoder}

\begin{table}[htbp]
\centering
\begin{tabular}{c|cc|cc}
\hline
\multirow{2}{*}{\textbf{Label}} & \multicolumn{2}{c|}{\textbf{QAE}} & \multicolumn{2}{c}{\textbf{SQC}} \\ \cline{2-5} 
               & \textbf{Avg}                 & \textbf{Std}                & \textbf{Avg}                  & \textbf{Std}                 \\ \hline
0              & 0.815                        & 0.080                       & \textbf{0.841}                & 0.073                        \\
1              & \textbf{0.700}               & 0.168                       & 0.679                         & 0.174                        \\
2              & 0.715                        & 0.115                       & \textbf{0.736}                & 0.117                        \\
3              & 0.699                        & 0.117                       & \textbf{0.725}                & 0.118                        \\
4              & 0.694                        & 0.093                       & \textbf{0.709}                & 0.116                        \\
5              & 0.705                        & 0.091                       & \textbf{0.706}                & 0.097                        \\
6              & 0.744                        & 0.082                       & \textbf{0.772}                & 0.091                        \\
7              & \textbf{0.703}               & 0.098                       & 0.689                         & 0.123                        \\
8              & 0.694                        & 0.088                       & \textbf{0.713}                & 0.093                        \\
9              & 0.633                        & 0.147                       & \textbf{0.671}                & 0.147                        \\ \hline
\end{tabular}
\caption{Comparative analysis of the QAE and SQC techniques applied to the Optical Recognition of Handwritten Digits dataset~\cite{Dua:2019, Alpaydin1998} under ideal simulation conditions. The columns labeled \textbf{Avg} and \textbf{Std} display the average fidelity and the respective standard deviation measured across 20 pairs of original and reconstructed states, offering insights into the effectiveness of each method in preserving quantum state information during compression. The bold values indicate the best result per label.
} \label{tab:comparison}
\end{table}

\begin{figure}[htbp]
    \centering
    \includegraphics[width=1.0\columnwidth]{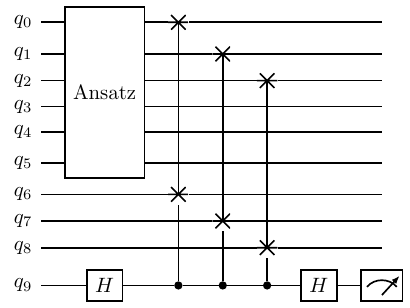}
    \caption{The figure displays the circuit employed to train the unitary operator, referred to as Ansatz, for the QAE. The first three qubits acted upon by the unitary are designated as trash qubits, whereas the last three are the latent qubits. The training process utilizes the SWAP Test with the goal of maximizing the fidelity between the trash state and the reference state. In this experiment, the reference state is $\ket{000}$, represented by qubits 6 to 8.}
    \label{fig:autoencoder}
\end{figure}

In this section, we contrast the performance of the Schmidt quantum compressor algorithm with that of the QAE, using fidelity as the benchmark, as expressed by Equation~\eqref{eq:fidelity}. Fidelity measures how closely the recovered state from the complete circuit (illustrated in Figure~\ref{fig:overview}) matches the initial quantum state. We anticipate that the fidelity outcomes from the Schmidt compressor algorithm will be comparable to those from the QAE, providing an important evaluation of the efficacy of these two quantum compression approaches.

The QAE ansatz, as depicted in Figure~\ref{fig:ansatz}, offers an efficient quantum circuit design. This architecture is considered hardware-efficient~\cite{kandala_hardware-efficient_2017,cerezo2020variational} because it employs only single-qubit $y$-rotations, represented as $R_y(\theta)$, along with CNOT gates between consecutive qubits. The rotation angles, denoted by $\theta$, are adjustable parameters that are fine-tuned through an optimization process. 

In our adaptation of the QAE methodology, we follow the general approach in which a classical computer is used to optimize a set of parameters, specifically the rotation angles for the $R_y(\theta)$ gates, within a parameterized quantum circuit. As illustrated in Figure~\ref{fig:autoencoder}, the quantum system is tasked with executing a SWAP test~\cite{barenco1996stabilisation} and carrying out measurements on an auxiliary qubit. These measurement results are then analyzed by a classical system to project and refine the model parameters using a learning algorithm. The objective is to enhance the fidelity between the subsystem B state, also known as the trash state, and a predefined reference state. In our case, the reference state is the computational basis state $\ket{000}$. This iterative cycle is repeated with the aim of improving towards the established goal.

This study employs the Optical Recognition of Handwritten Digits dataset~\cite{Dua:2019, Alpaydin1998}. The dataset was segmented into ten unique subsets, each representing one digit class in the range from 0 to 9. Every subset comprised 180 samples. For each digit class, 20 test samples were evaluated to calculate the mean fidelity and standard deviation. The simulation results are summarized in Table~\ref{tab:comparison}, which presents a comparative analysis of the QAE and the SQC based on fidelity metrics. Notably, the SQC outperformed the QAE in 8 out of 10 data compression tasks.

These findings highlight comparable fidelity levels between the two compression methods, with a slight advantage observed for the Schmidt compressor. The standout benefit of the Schmidt compressor is its independence from optimization, a process that can often be resource-intensive and time-consuming. Moreover, it circumvents issues commonly associated with optimization, such as barren plateaus, and eliminates the complex decision-making involved in selecting the optimal embedding and ansatz for variational circuits.

The simulations were conducted using the Qiskit Aer simulator~\cite{Qiskit}. For each single-digit class in the dataset, 20 samples were allocated to the test set and 160 to the training set. Prior to experimentation, the dataset features were standardized to a mean of 0 and a standard deviation of 1, and then rescaled to fit within the 0 to 1 range. The data vectors were subsequently normalized. The COBYLA optimization algorithm~\cite{Powell2007AVO} was utilized during the training phase, which concluded upon reaching 1000 iterations.

\subsection{Data classification}
\label{sec:classification}
A QAE can be utilized for classification by first compressing quantum data into a lower-dimensional quantum state, preserving the most significant features of the input data. This process makes the classifier more efficient by focusing on the most relevant information. After the compression, the encoded features, which capture the essential characteristics of the data, can be fed into a classifier (quantum or classical). This method allows the classifier to operate on a simplified yet informative representation, improving performance, especially in cases involving complex or high-dimensional data. The reduced dimensionality not only speeds up computation, but also potentially enhances the accuracy by eliminating noise and less relevant data from the inputs. Essentially, the QAE acts as a feature extractor in a machine learning pipeline, preparing data for subsequent classification stages.

Alternatively, the trash space---which represents the discarded aspects of the input data---can be analyzed to classify inputs based on the nature of the information lost during compression~\cite{bravo2021quantum,Park_2023_MLST,oh2023quantum}. This approach is particularly useful when the differences between classes can be captured by what is excluded from the primary model representation.

\begin{table}[htbp]
\centering
\begin{tabular}{c|c|c|c|c}
\hline
\textbf{$\phi$ (MCC)} & \multicolumn{2}{c|}{\textbf{SQC}} & \multicolumn{2}{c}{\textbf{QAE}} \\ \cline{1-5} 
\textbf{Label} & \textbf{Avg} & \textbf{Std} & \textbf{Avg} & \textbf{Std} \\ \hline
0 & {0.8850} & 0.0334 & \textbf{0.9797} & 0.0325 \\ 
1 & 0.4825 & 0.0141 & \textbf{0.6734} & 0.0679 \\ 
2 & {0.5622} & 0.0124 & \textbf{0.7689} & 0.1019 \\ 
3 & \textbf{0.6943} & 0.0217 & 0.6411 & 0.1047 \\ 
4 & \textbf{0.8341} & 0.0222 & 0.7561 & 0.0822 \\ 
5 & \textbf{0.6975} & 0.0230 & 0.6079 & 0.0810 \\ 
6 & {0.7927} & 0.0243 & \textbf{0.9254} & 0.0522 \\ 
7 & \textbf{0.7354} & 0.0156 & 0.6438 & 0.0783 \\ 
8 & {0.5296} & 0.0228 & \textbf{0.6228} & 0.0806 \\ 
9 & \textbf{0.5792} & 0.0148 & 0.5350 & 0.0861 \\ \hline
\end{tabular}
\caption{Analysis of the Schmidt quantum compression (SQC) and quantum autoencoder (QAE) applied in one-class classification for the Optical Recognition of Handwritten Digits dataset~\cite{Dua:2019, Alpaydin1998}. QAE requires optimization of 50 variational parameters, whereas SQC does not. A binary dataset was constructed for each class by distinguishing it from the remaining nine classes. The evaluation was conducted under ideal simulation conditions. The columns \textbf{Avg} and \textbf{Std} represent the average and the respective standard deviation of the Phi coefficient computed across 10 repetitions, respectively. The bold values indicate the best result per label.}
\label{tab:mcc}
\end{table}

Here, we employ the same methodology used for QAE, but substitute the parameterized compressor with the SQC combined with a classical neural network to evaluate the quantum circuit output. For each test sample $x_i$, the SQC is applied to extract real-valued information $y_i$ from the trash space, composed of 2 qubits. This is done through single-qubit state tomography of each qubit. The state tomography of the entire trash register would yield better results, but at the cost of an exponentially higher number of measurements. This information is then input into a simple single-layer neural network. The output is defined as $ g(y_i, w) = \sigma \left( w^{(0)} y_i + b^{(0)} \right) $. In this formula, $w^{(0)} \in \mathbb{R}^{1 \times 16}$ represents the weight matrix and $b^{(0)} \in \mathbb{R}^{1}$ is the bias of the network. The function $\sigma$ denotes the Sigmoid activation function. The Matthews Correlation Coefficient (MCC)~\cite{MATTHEWS1975442} is utilized to evaluate the performance of the classifier on the test samples, as presented in Table~\ref{tab:mcc}.

The Matthews correlation coefficient (MCC), commonly referred to as the Phi coefficient $\phi$~\cite{Yule1912}, is a statistical measure used to assess the strength of the association between two binary variables. It is conceptually similar to the Pearson correlation coefficient, but specifically designed for binary data. In the context of binary classification in machine learning, the Phi coefficient can be calculated directly from the elements of a confusion matrix: True Positives (TP), False Positives (FP), True Negatives (TN), and False Negatives (FN). The formula for the Phi coefficient is $\displaystyle \phi = \frac{TP \times TN - FP \times FN}{\sqrt{(TP+FP)(TP+FN)(TN+FP)(TN+FN)}} $. This coefficient ranges from -1 to 1, where 1 indicates perfect agreement, 0 no better than random chance, and -1 perfect disagreement.

The Phi coefficient is particularly useful for assessing the performance of a classifier as it provides a single measure that considers all four quadrants of the confusion matrix, making it more informative than simple accuracy, especially in cases of class imbalance. Unlike accuracy, which can be misleadingly high in datasets where one class significantly outnumbers the other, the Phi coefficient remains a reliable indicator because it accounts for both types of errors (FPs and FNs) as well as successes (TPs and TNs). Furthermore, it can be a better choice than the F1 score in scenarios where both types of classification errors (false positives and false negatives) are equally costly or when the dataset is highly imbalanced. The F1 score, while useful, primarily focuses on the positive class and balances the precision and recall, which might not fully capture the effectiveness of the classifier in dealing with the negative class, a gap that the Phi coefficient fills by evaluating the performance across all categories.

This application leverages the Optical Recognition of Handwritten Digits dataset. To ensure a balanced training process of the neural network, the dataset configuration includes 150 training samples for the correct label, each replicated 9 times. Each of the remaining nine classes is represented by 150 training samples as well. For testing purposes, we have allocated 20 samples per class, including the correct label. The Adam optimization algorithm~\cite{kingma2017adam} and Cross Entropy loss function were utilized during the training phase, with batches of 25 random samples and 1000 iterations.

The decomposition of the SQC unitaries used in this application consists of 62 CNOT gates and has a depth of 117 when transpiled with Qiskit using the basis gates $u$ (general single-qubit gates) and $cx$ (CNOT gates), without any circuit optimization.

We compared the one-class classification performance of the Schmidt Quantum Compression protocol with a QAE-based one-class classifier introduced in Ref.~\cite{Park_2023_MLST}. 
The QAE used in our study comprises 50 $R_y$ gates and 40 CZ gates, with a circuit depth of 33 when transpiled using Qibo~\cite{qibo}. Similar to SQC, we employed two trash qubits, with the total number of qubits is six. Due to the variational nature of QAE, we utilized 50 trainable parameters, corresponding to the number of $R_y$ gates. Each experiment consisted of 50 iterations with a batch size of ten to ensure convergence in the loss landscape, and the number of training and test samples was identical to that used in  SQC. 

The SQC demonstrated comparable performance to the variational quantum one-class classifier, highlighting its efficiency and potential advantages in machine learning applications.

\section{Optimization}
\label{sec:optimizations}

In the context of QAE, using a mixed state as a reference state helps address the limitations associated with using pure reference states for compressing quantum information~\cite{ma2023quantum}. Traditional QAEs often use pure states as reference states, which impose an upper bound on encoding fidelity, especially when dealing with high-rank states with high entropy. This upper bound restricts the compression rate and results in a low overlap between the trash state and the reference state.

\begin{table}[htbp]
\centering
\begin{tabular}{c|cc|cc|cc}
\hline
\multirow{2}{*}{\textbf{Label}} & \multicolumn{2}{c|}{\textbf{SQC}} & \multicolumn{2}{c|}{\textbf{Optim. 1}} & \multicolumn{2}{c}{\textbf{Optim. 2}}\\ \cline{2-7} 
                                & \textbf{Avg}    & \textbf{Std}    & \textbf{Avg}    & \textbf{Std}    & \textbf{Avg}    & \textbf{Std}    \\ \hline
0                               & 0.841           & 0.073           & \textbf{0.859}  & 0.060           & 0.851           & 0.066           \\
1                               & 0.679           & 0.174           & \textbf{0.794}  & 0.149           & 0.736           & 0.151           \\
2                               & 0.736           & 0.117           & \textbf{0.780}  & 0.100           & 0.750           & 0.110           \\
3                               & 0.725           & 0.118           & \textbf{0.740}  & 0.113           & 0.735           & 0.114           \\
4                               & 0.709           & 0.116           & \textbf{0.736}  & 0.109           & 0.722           & 0.114           \\
5                               & 0.706           & 0.097           & \textbf{0.730}  & 0.091           & 0.721           & 0.094           \\
6                               & 0.772           & 0.091           & \textbf{0.792}  & 0.084           & 0.781           & 0.090           \\
7                               & 0.689           & 0.123           & \textbf{0.718}  & 0.103           & 0.703           & 0.116           \\
8                               & 0.713           & 0.093           & \textbf{0.746}  & 0.086           & 0.738           & 0.086           \\
9                               & 0.671           & 0.147           & \textbf{0.703}  & 0.137           & 0.695           & 0.137           \\ \hline
\end{tabular}
\caption{Comparative analysis of the Schmidt Compression technique and its optimizations applied to the Optical Recognition of Handwritten Digits dataset~\cite{Dua:2019, Alpaydin1998} under ideal simulation conditions. The columns labeled \textbf{Avg} and \textbf{Std} represent the average fidelity and the corresponding standard deviation measured across 20 pairs of original and reconstructed states. Bold values indicate the best result for each label.
} \label{tab:comparison_optimization}
\end{table}

By allowing the reference state to be a mixed state, the entropy of the reference state can better match the entropy of the trash state, resulting in  higher decoding fidelity.

\begin{figure}[htbp]
    \centering
    \includegraphics[width=1.0\columnwidth]{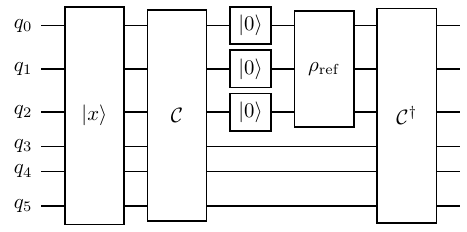}
    \caption{Schmidt Quantum Compression Protocol with Optimization 1. The complete trash state density matrix is estimated through tomography, and the most relevant eigenvector (corresponding to the largest eigenvalue) is initialized as the reference state.}
    \label{fig:optimization_1}
\end{figure}

This approach enhances the performance of QAEs by addressing the entropy inconsistency between initial and reconstructed states, thereby improving the fidelity of the compression and reconstruction processes.

Despite the advantages offered by using mixed states as the reference state, this technique is not efficient in terms of quantum operations. Initializing mixed states requires auxiliary qubits and a circuit depth that is exponential in the number of qubits in the reference system plus the number of auxiliary qubits~\cite{Ezzell_2023}.

Here, we present two efficient alternatives that offer quadratic and exponential improvements in the number of quantum operations, respectively, compared to using mixed states as proposed by~\cite{ma2023quantum}. These advantages are achieved through the use of classical data transmission.

Both optimizations use tomography to obtain information about the trash state. The first optimization, referred to as Optimization 1 and illustrated in Figure~\ref{fig:optimization_1}, performs tomography on the set of trash qubits to estimate the trash state for each instance of the compressor. The density matrix is decomposed via diagonalization, and only the most significant eigenvector (with the largest eigenvalue) is transmitted through classical communication and initialized in the reference space. Since the eigenvector is a pure state, it can be initialized in the reference state's amplitudes with a complexity of $O(2^{n_\textsc{b}})$, compared to $O(2^{n_\textsc{b}+n_\mathrm{aux}})$, where $n_\mathrm{aux}$ is the number of auxiliary qubits needed for initializing the mixed state. The selected eigenvector contains the most significant portion of the information of $\rho_t$. In our examples, $n_\textsc{b} = n_\mathrm{aux} = n/2$ (when the subsystems are of equal size), thus the complexity of the classical step under these conditions is $O(2^n)$, and the circuit complexity is $O(2^{n/2})$. The classically transmitted eigenvector has a length of $\sqrt{N}$.

\begin{figure}[htbp]
    \centering
    \includegraphics[width=1.0\columnwidth]{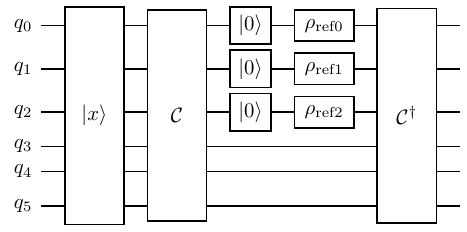}
    \caption{Schmidt Quantum Compression Protocol with Optimization 2. The density matrix of each qubit in the trash system is estimated through tomography, and the most relevant eigenvector (corresponding to the largest eigenvalue) is initialized in the qubits of the reference system.}
    \label{fig:optimization_2}
\end{figure}

Optimization 2, shown in Figure~\ref{fig:optimization_2}, has both classical and quantum costs that are linear in the number of trash qubits, $O(n_\textsc{b})$. In this case, tomography is performed qubit by qubit, producing $n_B$ density matrices of dimension $2 \times 2$. From each density matrix, only the eigenvector corresponding to the largest eigenvalue is transmitted classically. Consequently, we have $n_\textsc{b}$ single-qubit initializations in the reference system. Since the cost of single-qubit tomography is constant and the initialization of the single-qubit states corresponding to the eigenvectors also has a constant cost, the operation complexity (considering both classical and quantum steps) is $O(n_\textsc{b})$.

Table~\ref{tab:comparison_optimization} compares the results of the Schmidt compressor with and without optimizations. As expected, the optimizations achieve higher fidelities in all cases, with Optimization 1 yielding the best results. Optimization 2 proves to be a good compromise between cost and performance in this experiment. It is important to note that these optimizations are deterministic and do not rely on any machine learning or statistical techniques.

\section{Conclusion}
\label{sec:conc}
Quantum compression is a critical technique in quantum computing and data analysis, essential for optimizing quantum resources. In this study, we introduced the Schmidt Quantum Compressor, a deterministic quantum compression algorithm. This approach presents a viable alternative to variational quantum algorithms, such as QAEs, which require numerous repetitions due to their stochastic nature and computationally expensive parameter optimization. Our deterministic method circumvents several significant challenges associated with variational quantum algorithms, notably barren plateaus~\cite{mcclean_barren_2018, Cerezo2021bp, larocca2024review} which pose risks of converging to local minima or leading to the non-trainability of variational quantum circuits~\cite{thanasilp2022exponential}.

The Schmidt Quantum Compressor relies on identifying a typical state, $\ket{\psi}$, which serves as a reference for compression. In this study, the typical state is defined as the average state of the dataset, specifically chosen for its ability to minimize the distance between the dataset samples and $\ket{\psi}$, thereby optimizing the fidelity of the compression. By employing this approach, the Schmidt quantum compressor efficiently encodes quantum information with reduced computational overhead compared to traditional variational methods. It is capable of reconstructing quantum states with high fidelity, particularly when the states are close to $\ket{\psi}$. However, this method also introduces new challenges, especially when dealing with quantum data.

Our empirical studies, focusing on reconstruction fidelity using the handwritten digits dataset, 
demonstrate the practicality and effectiveness of the Schmidt Quantum Compressor. Numerical simulations revealed that our compression method consistently maintained high fidelity for the compressed states while operating within feasible error margins, making it competitive with QAE techniques. These findings reinforce the method's viability for real-world quantum system applications.

As future work, we propose utilizing data fusion techniques~\cite{lahat2015} to obtain the typical state, thereby overcoming the limitations of using the average as proposed in Section~\ref{sec:avg_state}; i.e., limited to real, non-negative, sample states. This approach has the potential to yield better results in the compression process for complex data. In particular, the High-Order Generalized Singular Value Decomposition (HO-GSVD)~\cite{kempf2023} can be applied to identify common attributes across multiple samples. This can be accomplished by tensorizing the samples, followed by appropriate processing to derive a vectorized representation of these shared attributes.

\section*{Data availability}
The data and software generated during this study are available at \url{https://github.com/qDNA-yonsei/qdna-examples} 
and \url{https://github.com/qDNA-yonsei/qdna-lib}~\cite{qdnalib}. 

\section*{Acknowledgments}
This research was supported by the Institute for Information \& communications Technology Promotion (IITP) grant funded by the Korean government (No. 2019-0-00003, Research and Development of Core Technologies for Programming, Running, Implementing and Validating of Fault-Tolerant Quantum Computing System), the Yonsei University Research Fund of 2024 (2024-22-0147), and the National Research Foundation of Korea (Grant Numbers: 2022M3E4A1074591, 2023M3K5A1094805, 2023M3K5A1094813), the KIST Institutional Program (2E32941-24-008), and CNPq/MCTI (No. 350419/2024-8, Multiplexed quantum communication for hybrid quantum networks). N.A.R.B. acknowledges funding from the European Research Council (Consolidator Grant ‘Cocoquest’ 101043705) and the Austrian Research Promotion Agency (FFG) through the project FO999914030 (MUSIQ), funded by the European Union – NextGenerationEU. N.A.R.B. also acknowledges support from the Miller Institute for Basic Research in Science at the University of California, Berkeley, during the initial months of the project. 

\section*{Conflict of interests}

All authors declare no conflicts of interest.

\bibliography{references}

\end{document}